# Low-frequency noise parameter extraction method for single layer graphene FETs

Nikolaos Mavredakis, Wei Wei, Emiliano Pallecchi, Dominique Vignaud, Henri Happy, Ramon Garcia Cortadella, Nathan Schaefer, Andrea Bonaccini Calia, Jose Antonio Garrido, and David Jimenez

*Abstract*—in this paper, a detailed parameter extraction methodology is proposed for low-frequency noise (LFN) in single layer (SL) graphene transistors (GFETs) based on a recently established compact LFN model. Drain current and LFN of two short channel back-gated GFETs (L=300, 100 nm) were measured at lower and higher drain voltages, for a wide range of gate voltages covering the region away from charge neutrality point (CNP) up to CNP at p-type operation region. Current-voltage (IV) and LFN data were also available from a long channel SL top solution-gated (SG) GFET (L=5 μm), for both p- and n-type regions near and away CNP. At each of these regimes, the appropriate IV and LFN parameters can be accurately extracted. Regarding LFN, mobility fluctuation effect is dominant at CNP and from there the Hooge parameter $\alpha_H$ can be extracted while the carrier number fluctuation contribution which is responsible for the well-known M-shape bias dependence of output noise divided by squared drain current, also observed in our data, makes possible the extraction of the $N_T$ parameter related to the number of traps. In the less possible case of a Λ-shape trend, $N_T$ and $\alpha_H$ can be extracted simultaneously from the region near CNP. Away from CNP, contact resistance can have a significant contribution to LFN and from there the relevant parameter $S_{\Delta R}^2$ is defined. The LFN parameters described above can be estimated from the low drain voltage region of operation where the effect of Velocity Saturation (VS) mechanism is negligible. VS effect results in the reduction of LFN at higher drain voltages and from there the IV parameter $h\Omega$ which represents the phonon energy and is related to VS effect can be derived both from drain current and LFN data.

*Index Terms*— Graphene transistor (GFET), single-layer, low-frequency noise, compact model, parameter extraction

## I. INTRODUCTION

GRAPHENE devices (GFETs) have been given considerable attention over the last years since they can achieve very high speed performance [1]-[2] which makes them an ideal candidate for future RF applications. In this kind of applications though, the effect of low-frequency noise (LFN) cannot be neglected since it can deteriorate their efficiency. More specifically, LFN can be up-converted to phase-noise and degrade the performance of voltage control [3] or ring [4] oscillators as well as terahertz detectors [5-6]. In addition, it can reduce the sensitivity of chemical or biological sensors [7]-[9].

Because of the aforementioned effects of LFN on state of the art GFET applications, thorough research has been conducted recently [10]-[19] which has contributed to the comprehension of the main mechanisms that generate LFN in these devices. While several compact models have been established, simpler [12]-[16] or more physics-based [18]-[19], which can be easily integrated in circuit simulators, there is no complete methodology proposed for the extraction of LFN parameters in GFETs. Procedures for the extraction of some Current-Voltage (IV) parameters such as charge neutrality point (CNP) voltage $V_{CNP}$, mobility $\mu$, residual charge $\rho_0$ and contact resistance $R_c$, [17], [20]-[21] are already there and they can be proved very helpful to the calculation of LFN parameters.

The methodology for the extraction of LFN parameters proposed in this work is based on a complete chemical-potential based compact GFET model regarding CV-IV [22]-[23], RF [24] and LFN [18]-[19] behavior. The schematics of the devices under test are illustrated in Fig. 1a, 1b where the electrostatics can be described by the equivalent capacitive circuit in Fig. 1c [18]-[19], [22]-[24]. There $C_{top}$ and $C_{back}$ are the top and back gate oxide capacitances and $C_q$ represents the quantum capacitance where the graphene charge $Q_{gr}$ is stored. The voltage drop across $C_q$ equals to the chemical potential $V_c$ which is defined as the potential difference between the quasi-Fermi level and CNP and is equivalent to the surface potential of MOSFETs. A linear dependence is considered between $C_q$ and $V_c$ ($C_q(C_q=k V_c)$ where coefficient $k$ is defined in [22].

The LFN model proposed in [18]-[19] shows that the three main effects that generate LFN in GFETs are: carrier number fluctuation ($\Delta N$), mobility fluctuation ($\Delta\mu$) and contact resistance contribution ($\Delta R$), similarly as in most semiconductor devices. $\Delta N$ effect which is responsible for the M-shape gate-bias dependence of output LFN divided by squared drain current ($S_{ID}/I_D^2$) with a minimum at CNP [18], is the result of trapping/detrapping mechanism near the dielectric interface of the device [25]-[27]. $\Delta N$ LFN models for MOSFETs in [26], [28]-[31] propose an $\sim (g_m/I_D)^2$ dependence of $S_{ID}/I_D^2$ which is a useful approximation but valid only under uniform channel conditions at low drain voltage regime, which

N. Mavredakis and D. Jimenez are with the Departament d'Enginyeria Electrònica, Escola d'Enginyeria, Universitat Autònoma de Barcelona, Bellaterra 08193, Spain. (e-mail: Nikolaos.mavredakis@uab.es).

W. Wei, E. Pallecchi, D. Vignaud and H. Happy are with Univ. Lille, CNRS, UMR 8520 - IEMN, F-59000 Lille, France.

R. G. Cortadella, Nathan Schaefer, A. Bonaccini C. and J. A. Garrido are with the Catalan Institute of Nanoscience and Nanotechnology (ICN2), CSIC, Barcelona Institute of Science and Technology, Campus UAB, Bellaterra, Barcelona, Spain.

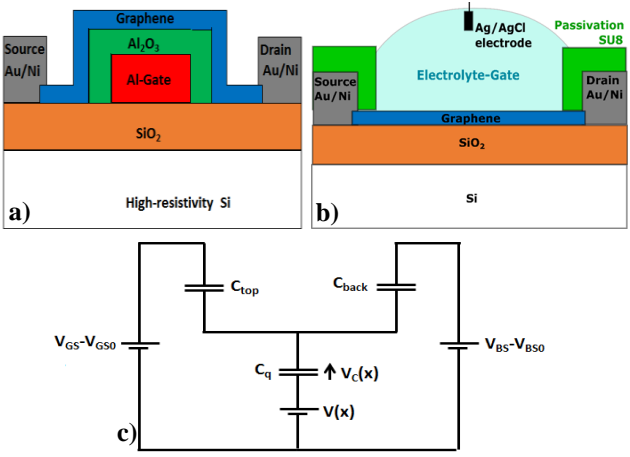

Fig. 1. Schematic of the a) back-gated Al₂O₃ single-layer GFET and b) top solution-gated single-layer GFET. c) Equivalent capacitive circuit.

has also been applied in GFETs [12]-[13]. The model proposed in [18]-[19] and used in the current study is a complete model valid in all regions of operations since it takes into account the non-homogeneity of the device channel at higher drain voltage conditions [18], [32]-[33]. $\Delta N$ effect can also provide a Λ-shape in case of a small $\rho_0$ value [18]. In fact, for GFETs on h-BN substrates [16], $\rho_0$ value becomes very low and as a result not only LFN is reduced but also M-shape is eliminated since the higher the $\rho_0$ the more intense the M-shape dependence of LFN [18]. A direct comparison of our LFN model with data from [16] is shown in [18, figure 5b] with precise results. Moreover, number of traps and consequently LFN is shown to be reduced after electron-beam irradiation of GFETs in [15] where a simple $\Delta N$ LFN model is also used which is incapable of capturing the LFN data. The specific model only accounts for the $\sim 1/Q_{gr}^2$ dependence of LFN and neglects the $\delta g_{gr}/\delta Q_{tr}$ ratio which is really significant [18, equations 1, 3-4], [26] and thus it is reliable only away CNP while our proposed LFN model [18]-[19] is valid in all regions of operation. $\Delta \mu$ effect is caused by the fluctuations of the bulk mobility and is empirically characterized by Hooge formula [34]. Regarding its gate-bias dependence, $S_{ID}/I_D^2$ $\Delta \mu$ LFN always gives a Λ-shape with a maximum at CNP [18]. $\Delta \mu$ LFN or volume noise dominates when the number of graphene layers is high while the lower the number of the layers the higher is the contribution of $\Delta N$ LFN or surface noise [14]. In this work as well as in [18]-[19], single-layer (SL) GFETs are studied and thus, $\Delta N$ LFN prevails but $\Delta \mu$ LFN also plays a role near CNP. Increased $R_c$ values especially at short channel GFETs, can contribute importantly to LFN away from CNP while short channel effects like Velocity Saturation (VS) reduce LFN in GFETs mainly at high electric fields [19].

The proposed parameter extraction methodology in this work is firstly applied to experimental data from back-gated Al₂O₃ SL GFETs (Fig. 1a) where graphene grown by CVD on a copper foil was used [35]-[38]. Drain current and LFN were measured on two short channel GFETs with $W=12$ $\mu m$ and $L=300$ nm (C300) and $L=100$ nm (D100) respectively, for gate voltages ($V_G$) from far away CNP at p-type region up to CNP and the vicinity of n-type region, at low and high drain voltage values ($V_{DS}=$ 60, 100, 200 mV for C300 and $V_{DS}=$ 30, 60, 300 mV for D100). $V_G$ sweep was not extended further to n-type regime due to asymmetry of both IV and LFN data there [19]. The method is also used at a SL solution-gated graphene transistor (SG GFET) (Fig. 1b) with $W=40$ $\mu m$ and $L=5$ $\mu m$ [39]. IV and LFN data were measured from strong p-type to strong n-type region only at low drain voltage ($V_{DS}=40$ mV). Since the latter is a long device and the $V_{DS}$ is low, VS related parameters cannot be extracted since they are significant only at high electric field conditions. Besides for the specific SG-GFET, $R_c$ is quite low, as it will be shown, and hence does not affect LFN. In order to study the bias dependence of normalized $S_{ID}/I_D^2$ LFN, its value at $1$ $Hz$ was calculated by averaging from $10$ $Hz$ to $40$ $Hz$. For more details on fabrication and measurements procedure see [19], [35]-[39].

The most fundamental CV-IV parameters ($C=C_{top}+C_{back}$, $V_{CNP}$, $\mu$, $\rho_0$, and $R_c$) should be extracted accurately since they are used in LFN equations [18]-[19]. Thus, initially a well-established procedure for the extraction of the above parameters was applied [17], [20]-[21] and then the estimated values were used for the simulations of the drain current Verilog-A model [22]-[24]. The validation of the model with experimental $I_D$ data for each device under test (DUT) and for all regions of operation is accurate as depicted in transfer characteristics of Fig. 2 vs. gate voltage overdrive $V_{GEFF}=V_G-V_{CNP}$. C300 and D100 GFETs are shown in Fig. 2a, 2b where left subplots correspond to low $V_{DS}$ and right subplots to higher $V_{DS}$. SG-GFET is shown in Fig. 2c.

## II. LFN MODELLING

A new LFN model was recently proposed for SL GFETs describing the bias dependence of LFN including also the VS effect contribution to it. The normalized output noise $S_{ID}/I_D^2$ is given by the sum of $\Delta N$, $\Delta \mu$ and $\Delta R$ contributions [18]-[19]:

$$\frac{S_{I_D}}{I_D^2} = \frac{S_{I_D}}{I_D^2}\bigg|_{\Delta N} + \frac{S_{I_D}}{I_D^2}\bigg|_{\Delta \mu} + \frac{S_{I_D}}{I_D^2}\bigg|_{\Delta R} =$$

$$\frac{S_{I_D}}{I_D^2}\bigg|_{\Delta NA} - \frac{S_{I_D}}{I_D^2}\bigg|_{\Delta NB} + \frac{S_{I_D}}{I_D^2}\bigg|_{\Delta \mu A} - \frac{S_{I_D}}{I_D^2}\bigg|_{\Delta \mu B} + \frac{S_{I_D}}{I_D^2}\bigg|_{\Delta R} \quad (1)$$

$\Delta N$ LFN contribution is given by [18]-[19]:

$$\frac{S_{I_D}}{I_D^2}\bigg|_{\Delta NA} = \frac{A1}{2k\left(\alpha k + C^2\right)g_{vc}}$$

$$\left[-2\sqrt{\alpha k}\,C \arctan\left(\sqrt{\frac{k}{\alpha}}V_c\right) \pm \alpha k \ln\left(\alpha + kV_c^2\right) \pm 2C^2 \ln\left(C \pm kV_c\right)\right]_{V_{cd}}^{V_{cs}} \quad (2)$$

$$\frac{S_{I_D}}{I_D^2}\bigg|_{\Delta NB\,[V_c] \le V_{ccrit}} = \frac{B1}{2\left(\alpha k + C^2\right)^3 S}$$

$$\left[\mp\frac{2C^3 k\left(C^2 + ak\right)}{C \pm kV_c} \mp \frac{k\left(C^2 + ak\right)\left(3aC^2 + a^2k \pm 2C^3V_c\right)}{\alpha + kV_c^2}\right.$$

$$\left.+\frac{2C^3\sqrt{k}\left(C^2 - 3ak\right)}{\sqrt{\alpha}} \arctan\left(\sqrt{\frac{k}{\alpha}}V_c\right) \pm k\left(-3C^4 + aC^2k\right)\ln\left[\frac{a + kV_c^2}{\left(C \pm kV_c\right)^2}\right]\right]_{V_{cd}}^{V_{cs}} \quad (3a)$$

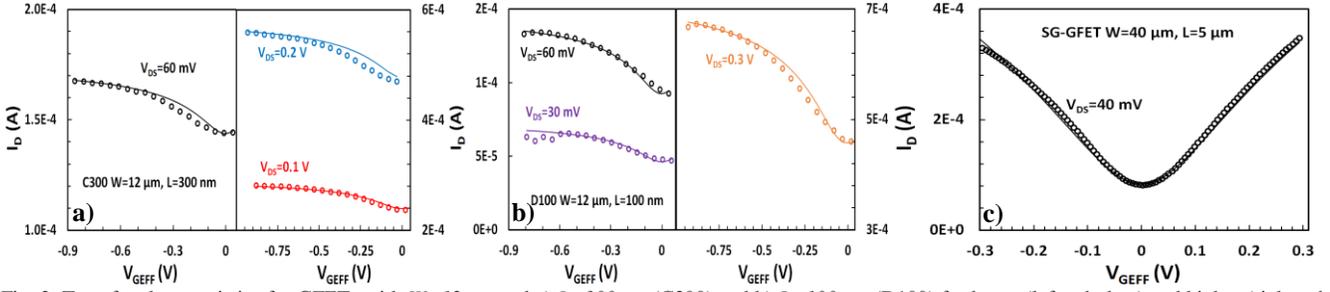

Fig. 2. Transfer characteristics for GFETs with $W=12\ \mu m$ and a) $L=300\ nm$ (C300) and b) $L=100\ nm$ (D100) for lower (left subplots) and higher (right subplots) available $V_{DS}$ values. c) Transfer characteristic for SG-GFET with $W=40\ \mu m$ and $L=5\ \mu m$. Measurement (markers), model (lines).

$$\left.\frac{S_{I_D}}{I_D^2}\right|_{\Delta NB_{|V_c|\le V_{ccrit}}} = \frac{B1}{N}$$

$$\left.\frac{S_{I_D}}{I_D^2}\right|_{\Delta NB_{|V_c|>V_{ccrit}}} =$$

$$\left[\mp k\frac{\sqrt{V_c^2+\frac{\alpha}{k}}\left[a^2k(\pm c+kV_c)+C^3V_c(2C+3kV_c)+aC^2(\pm 4C+3kV_c)\right]}{\left(C^2+ak\right)^2(\pm C+kV_c)\left(\alpha+kV_c^2\right)}\pm\right.$$
$$\left. C^2\left(2C^2-ak\right)\ln\left[\frac{C\pm kV_c}{a\mp CV_c+\sqrt{C^2+\alpha k}\sqrt{V_c^2+\frac{\alpha}{k}}}\right]\right]_{V_{cd}}^{V_{cs}}$$
$$\frac{}{\left(C^2+ak\right)^{5/2}}$$
(3b)

with $A1=\frac{4KT\lambda N_T}{CWL_{eff}^2}$, $B1=\frac{A1\mu}{L_{eff}^2}$ where $N_T$ is the oxide volumetric trap density per unit energy ($ev^{-1}cm^{-3}$) used as an LFN model parameter, $K$ is the Boltzman constant, $T$ is the temperature, $\lambda=0.1\ nm$ is the tunneling attenuation distance, $V_{cs,d}$ is the chemical potential at source and drain side respectively, $gvc$ is a normalized drain current coefficient [18], [22] and $\alpha=2e$ $\rho_0$ is a residual charge related term [18] where $e$ is the electron charge. $W$ is the width of the device while $L_{eff}$ is an effective length because of the VS effect [19], [22]. For the later, a two-branch model is used [19, equation 4] below and above a critical value of graphene net charge, $Q_{ccrit}=e\Omega^2/2\pi u_f$ which designates two regions of operation, near and away CNP respectively; $u_f$ ($\sim 10^6\ m/s$) is the Fermi velocity. Saturation velocity is constant $u_{sat}=S=2u_f/\pi$ for $|V_c|<V_{ccrit}$ while for $|V_c|>V_{ccrit}$, $u_{sat}=N/\sqrt{(V_c^2+\alpha/\kappa)}$ with $N=h\Omega u_f/e$; $h\Omega$ is the phonon energy and is used as an IV model parameter where $h$ is the Planck constant and $\Omega$ the frequency of the radiation while $V_{ccrit}$ can be calculated directly from $Q_{ccrit}$ [19]. Equation (2) represents the long channel $\Delta N$ LFN while equations (3a-3b) its reduction caused by VS effect [19]. $\Delta\mu$ LFN effect is calculated by [18]-[19]:

$$\left.\frac{S_{I_D}}{I_D^2}\right|_{\Delta\mu A} = \frac{A2}{g_{vc}}\left[CV_c\pm\frac{kV_c^2}{2}\right]_{V_{cd}}^{V_{cs}}$$
(4)

$$\left.\frac{S_{I_D}}{I_D^2}\right|_{\Delta\mu B_{|V_c|\le V_{ccrit}}} = \frac{B2k}{S}\left[\frac{2C}{\sqrt{\alpha k}}\arctan\left(\sqrt{\frac{k}{\alpha}}V_c\right)\pm 0.5\ln\left(a+kV_c^2\right)\right]_{V_{cd}}^{V_{cs}}$$
(5a)

$$\left.\frac{S_{I_D}}{I_D^2}\right|_{\Delta\mu B_{|V_c|>V_{ccrit}}} = \frac{B2}{N}\left[\pm k\sqrt{V_c^2+\frac{\alpha}{k}}+2C\ln\left(V_c+\sqrt{V_c^2+\frac{\alpha}{k}}\right)\right]_{V_{cd}}^{V_{cs}}$$
(5b)

with $A2=\frac{2a_He}{\kappa CWL_{eff}}$, $B2=\frac{A2\mu}{L_{eff}}$ where $\alpha_H$ is the unitless Hooge model parameter. Similarly to $\Delta N$ case, equation (4) derives the long channel $\Delta\mu$ LFN while equations (5a-5b) account for its reduction due to VS mechanism [19]. The way that equations (2-5) are solved depending on the signs of $V_{cs,d}$ and if their absolute value is lower or higher than $V_{ccrit}$ can be found in [19]. Finally the $\Delta R$ LFN [32] is given by [19]:

$$\left.\frac{S_{I_D}}{I_D^2}\right|_{\Delta R} = \frac{g_{ms}^2+g_{md}^2}{\left[1+\frac{R_C}{2}\left(g_{ms}+g_{md}\right)\right]^2}S_{\Delta R^2}, \quad g_{ms,d}=\frac{\mu Wk}{2L_{eff}}\frac{C_{top,back}}{C}V_{cs,d}^2$$
(6)

where $S_{\Delta R^2}$ is a model parameter ($\Omega^2/Hz$) and $g_{ms,d}$ are the source and drain transconductances, respectively [19], [23].

## III. PARAMETER EXTRACTION METHODOLOGY

The parameter extraction procedure developed in this paper is analyzed in the diagram of Fig. 3. In steps 1 and 2 of this chart, the derivation of the CV-IV parameters that are essential for the LFN parameters' extraction is described. Firstly, interface capacitances can be calculated by CV data and in the case of SG-GFET, $C_{top}=C\approx 2.13\ \mu F/cm^2$ [39]. On the other hand for the C300, D100 $Al_2O_3$ devices, CV data are not available, thus aluminum oxide back gate capacitance $C_{back}$ can be calculated as: $C_{back}\approx C=\varepsilon_{Al2O3}\varepsilon_0/t_{Al2O3}$ where $\varepsilon_{Al2O3}\varepsilon_0$, $t_{Al2O3}$ are the relative permittivity and thickness of the $Al_2O_3$ [19], [35]-[38], respectively. In order to calculate the rest of IV parameters, the total resistance is estimated from drain current data as: $R_{tot}=V_{DS}/I_{Ddata}$ at the lowest $V_{DS}$ available (C300: $V_{DS}=60\ mV$, D100: $V_{DS}=30\ mV$ and SG-GFET: $V_{DS}=40\ mV$). $R_{tot}$ is then illustrated in Fig.4a-4c for each DUT respectively. Markers represent the measurements whereas the voltage at CNP can be calculated as the value of $V_G$ at which $R_{tot}$ gets maximum. $V_{GEFF}=V_G$-$V_{CNP}$ is then calculated and used as x-axis in Fig. 4. After that, contact resistance $R_C$ can be extracted by using an extrapolation method [17]. Insets in Fig. 4a-4c illustrate $R_{TOT}$ vs. $1/V_{GEFF}$ far away from CNP. If this dependence is extrapolated to $1/V_{GEFF}=0$ then a linear fit can reliably provide $R_C$ for each of the DUTs. The next step is to calculate the mobility by [17]:

$$\mu = \frac{L}{R_{eff}C_gV_{GEFF}W}$$
(7)

where $C_g\approx C$ since $\mu$ is derived quite away from CNP:

$$R_{eff}=\frac{R_{tot}-R_C}{1-\sigma_0(R_{tot}-R_C)}, \quad \sigma_0=\sigma(CNP)$$
(8)

where $\sigma_0$ is the conductivity $\sigma=L/[W(R_{tot}-R_c)]$ at CNP. Then $\rho_0$ can be derived as [20]-[21]:

$*\pm,\mp$ : Top sign refers to $V_c>0$ and bottom sign to $V_c<0$.

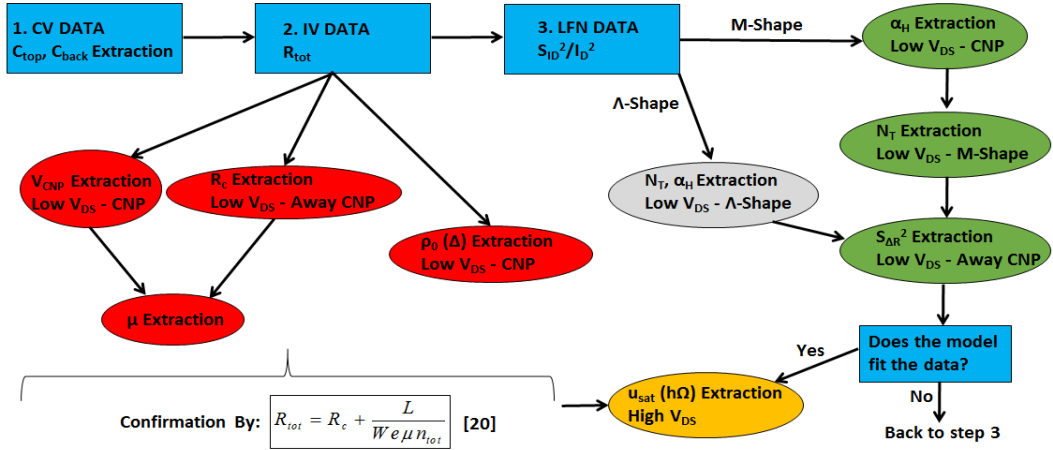

Fig. 3. IV and LFN parameter extraction flow chart.

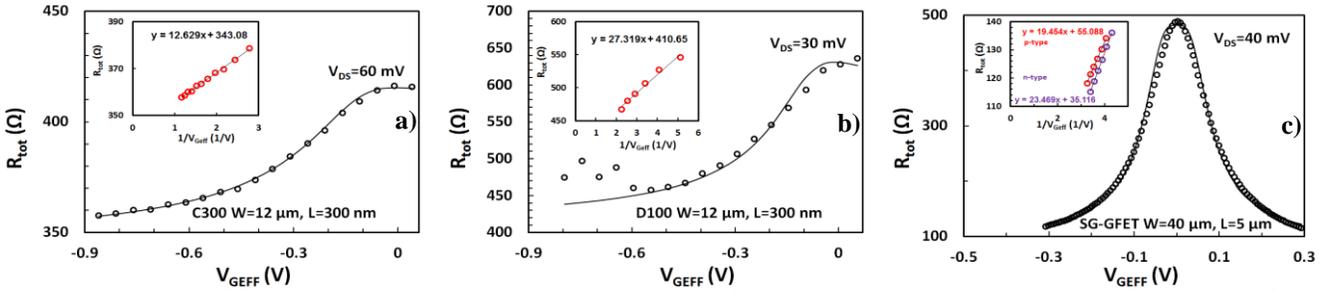

Fig. 4. Total resistance $R_{tot}$ vs gate voltage overdrive $V_{GEFF}$ for a) C300 at $V_{DS}=60\ mV$, b) D100 GFET at $V_{DS}=30\ mV$ and c) SG-GFET at $V_{DS}=40\ mV$. IV model parameters are extracted by the measurements (markers) and afterwards used in simulations (lines). The insets show the contact resistance extraction method with extrapolation of linear fitted line y at 0.

$$n_{tot} = \sqrt{n^2 + \frac{\rho_0^2}{e^2}}, \quad \sigma = n_{tot} e\mu \Rightarrow \rho_0 = \frac{\sigma_0}{e\mu} \qquad (9)$$

where $n_{tot}$, $n$ are the total and net densities respectively. In the compact model [22]-[23], parameter $\Delta$ expressing the inhomogeneity of the electrostatic potential is used which is related to $\rho_0$ as: $\rho_0 = \Delta/(\pi h^2 u_f)$ and thus, $\Delta$ can be estimated. After all the above parameters are extracted, theoretical $R_{tot}$ can be calculated [20]:

$$R_{tot} = R_c + R_{ch} = R_c + \frac{L}{W\sigma} = R_c + \frac{L}{Wn_{tot}e\mu} \qquad (10)$$

with $n$ estimated from [20]:

$$V_{GEFF} = \frac{en}{C} + \frac{hu_f\sqrt{n\pi}}{e} \qquad (11)$$

Solid lines in Fig.4 correspond to equation (10) and confirm the correct extraction of the IV parameters described above. The $h\Omega$ parameter related to VS effect, can be estimated at high $V_{DS}$ regime for the C300, D100 GFETs as shown in right subplots of Fig. 2a, 2b. Step 3 of Fig. 3 presents the LFN parameters' extraction methodology. Since LFN expressions in equations (2-5) are expressed through $V_{cs,d}$, the later should be calculated as in [22]-[23]. Internal $V_{Din}$, $V_{Sin}$ due to the voltage drop at source and drain because of $R_c$ are required for $V_{cs,d}$ calculation and can be estimated from: $V_{S,Din} = V_{S,D} - I_{Ddata}R_c/2$. Coefficient $gvc$ can also be derived if $V_{cs,d}$ and $V_{S,Din}$ are known [22]. In case of an M-shape trend of $S_{ID}/I_D^2$ LFN which is the case in our data and in most experimental findings in bibliography [12]-[19], the right branch of step 3 in Fig. 3 is followed where three sub-steps are required for the extraction of LFN parameters at low $V_{DS}$ region. Fig. 5 depicts C300 GFET in the upper plots and D100 GFET in the bottom ones where $S_{ID}/I_D^2$ LFN is shown in left and center column plots, at *1 Hz*, vs. $V_{GEFF}$; markers represent the data and lines the model and its different contributions. Each plot in the left and center columns contains two subplots in each of which one step of the LFN parameter extraction process is illustrated. The three first subplots of C300 and D100 GFETs describe the step $N_T$, $\alpha_H$ and $S_{\Delta R}^2$ parameters are extracted at low $V_{DS}$ while the right subplots of center column plots illustrate the case of high $V_{DS}$ where $h\Omega$ parameter can be extracted and compared to the value derived from IV data, as analyzed before. Regarding the SG-GFET LFN parameter extraction method, $S_{ID}/I_D^2$ LFN is shown in left and center plots of Fig. 6, at *1 Hz*, vs. $V_{GEFF}$ similarly as C300 and D100 GFETs as described above. Again, markers represent the data and lines the model and its different contributions. LFN data asymmetry at SG-GFET makes it essential to extract different LFN parameters for p-(left subplots) and n-type (right subplots) region. In more detail, initially $\alpha_H$ parameter of $\Delta\mu$ LFN can be extracted by fitting equation (4) with the minimum of M-shape observed at CNP as shown in Fig.5 (left subplots of left plots) and Fig. 6 (left subplots of left and center plots) since $\Delta\mu$ effect is dominant there at low $V_{DS}$. Then $N_T$ parameter of $\Delta N$ LFN can be calculated by fitting equation (2) with the maximum of M-shape near CNP according to Fig.5 (right subplots of left plots) and Fig. 6 (right subplots of left and center plots). It should be mentioned that equations (3, 5) are negligible at low $V_{DS}$ region [19]. In the case of a $\Lambda$-shape trend of $S_{ID}/I_D^2$ LFN, which is something not very usual, left branch of step 3 in Fig. 3 is followed. $\Delta N$ and $\Delta\mu$ models should be initially

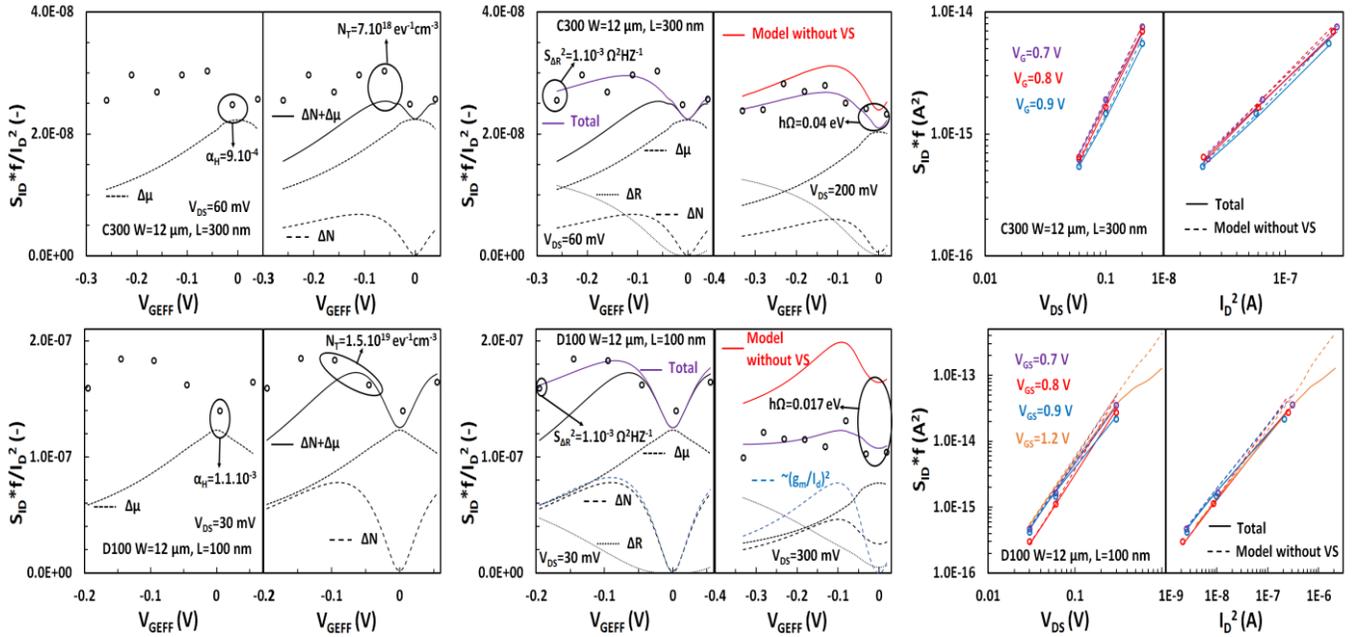

Fig. 5. Drain current noise divided by squared drain current and drain current noise $S_{ID}/I_D^2$ and $S_{ID}f$ respectively, both referred to *1 Hz*, for C300 GFET in upper plots and D100 GFET in bottom plots. $S_{ID}/I_D^2$ is shown at left and center column plots vs. gate voltage overdrive $V_{GEFF}$, where LFN parameter extraction method is shown in analytical steps for low $V_{DS}$ ($V_{DS}=60$ mV for C300, $V_{DS}=30$ mV for D100) at first column plot and left subplot of center column plot and high $V_{DS}$ ($V_{DS}=200$ mV for C300 and $V_{DS}=300$ mV for D100) in right subplot of second column plot. $S_{ID}f$ is shown at right column plots vs. drain voltage $V_{DS}$ at left subplots and vs. $I_D^2$ at right subplots for $V_{GS}=0.7$, $0.8$, and $0.9$ V; for D100 device, model is also shown for $V_{GS}=$ *1.2 V* and for $V_{DS}$ up to *1 V*. Measurement (markers), total model [19] (solid purple lines in center column plots and solid lines in right plots), $\Delta N+\Delta\mu$ sum (solid black lines in center column plots), model without VS effect [18] (solid red lines in center plots and dashed lines in right plots), different noise contributions $\Delta N$, $\Delta\mu$, $\Delta R$ (dashed and dotted lines respectively). Simplified ~ $(g_m/I_D)^2$ noise model is shown with blue lines for D100 device (blue dashed line in center bottom plot) for comparison reasons.

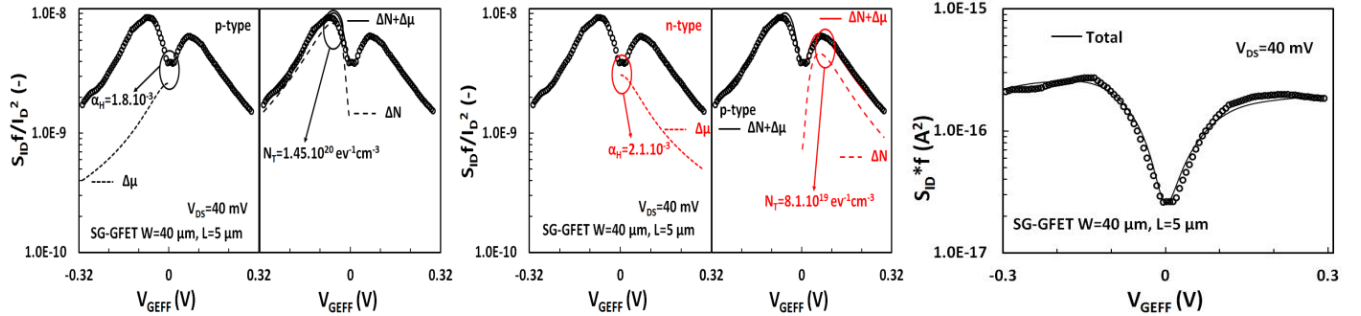

Fig. 6. Drain current noise divided by squared drain current and drain current noise $S_{ID}/I_D^2$ and $S_{ID}f$ respectively, both referred to *1 Hz*, vs. gate voltage overdrive $V_{GEFF}$ for SG-GFETs. $S_{ID}/I_D^2$ is shown at left and center plots where LFN parameter extraction method is shown in analytical steps for low $V_{DS}$ ($V_{DS}=40$ mV) while $S_{ID}f$ is shown at right column plot. Measurement (markers), total model [19] (solid black lines; solid red line in right subplot of center plot for n-type region), different noise contributions $\Delta N$, $\Delta\mu$ (dashed and dotted lines respectively). In eq. l, $\Delta N+\Delta\mu$ sum equals to total model since $\Delta R$ is negligible for SG-GFET.

tested separately to check if they fit the data. If either of them works then the relevant parameter can be extracted ($\alpha_H$ or $N_T$). On the contrary, if none of them work alone then both parameters contribute simultaneously and they have to be extracted carefully by trying to fit the maximum and the slope of $\Lambda$-shape near CNP. Finally, an increase of $S_{ID}/I_D^2$ far away from CNP indicates the contribution of $R_C$ to LFN, and from there the $S_{\Delta R}^2$ parameter of $\Delta R$ LFN can be estimated with equation (6) as it is shown in Fig.5 (left subplots of center plots). If the first iteration of the estimation of LFN parameters does not give good model fitting in general, then the procedure is repeated until the model is precise. As shown in [19], VS effect reduces LFN for high $V_{DS}$ values and this is experimentally confirmed in Fig. 5 (right subplots of center plots) for C300, D100 GFETs, respectively. By applying the $h\Omega$ parameter value that has already been extracted from IV data, at equations (3, 5), the

LFN model fits well the experiments while the LFN model without VS is also shown for comparison reasons with red solid lines. In the SG-GFET, $h\Omega$ parameter cannot be extracted since VS is negligible under low drain voltage and long channel length conditions while $R_c$ is also too low to affect LFN. Table I shows all the IV and LFN parameters extracted for each DUT. Finally, ~ $(g_m/I_D)^2$ $\Delta N$ LFN model [26], [28]-[31] is also shown for D100 GFET in center-bottom plot of Fig. 5 with blue dashed line. In left subplot where $V_{DS}$ is low and consequently channel is uniform, ~ $(g_m/I_D)^2$ $\Delta N$ LFN approach agrees with our model but at right subplot the high $V_{DS}$ breaks the homogeneity of the channel and as a result ~ $(g_m/I_D)^2$ $\Delta N$ LFN model is not accurate [18], [33]. The fact that ~ $(g_m/I_D)^2$ $\Delta N$ LFN is equivalent to our $\Delta N$ model for the same $N_T$ parameter confirms the validity of both models under low $V_{DS}$ conditions and consequently the rest of LFN parameters ($\alpha_H$ and $S_{\Delta R}^2$) are extracted correctly there. If the above parameters ($N_T$, $\alpha_H$ and $S_{\Delta R}{}^2$) are re-adjusted to fit the total LFN model with data for high $V_{DS}$ when $\sim (g_m/I_D)^2 \; \Delta N$ LFN model is used (Total = $\sim (g_m/I_D)^2 \; \Delta N + \Delta\mu + \Delta R$), then the accuracy achieved at low $V_{DS}$ region will be lost and this proves the validity of our model vs. $\sim (g_m/I_D)^2 \; \Delta N$ LFN one; $h\Omega$ cannot be re-adjusted since it is correctly extracted from IV data.

TABLE I
IV-LFN Model extracted Parameters

| Parameter | Units | C300 | D100 | SG-GFET | |
|---|---|---|---|---|---|
| $\mu$ | cm²/(V·s) | 2000 | 300 | 5450 | |
| $C_{ox}$ | μF/cm² | 1,35 | 1,35 | 2,13 | |
| $V_{CNP}$ | V | 0,83 | 0,78 | 0,187 | |
| $\Delta$ | eV | 0.116 | 0.092 | 0,051 | |
| $h\Omega$ | eV | 0.04 | 0.017 | - | |
| $R_s/2 = R_{SD}$ | Ω | 172 | 205 | 23 | |
| | | | | **p-type** | **n-type** |
| $N_T$ | eV⁻¹cm⁻³ | $7 \cdot 10^{18}$ | $1.5 \cdot 10^{19}$ | $1.45 \cdot 10^{20}$ | $8.1 \cdot 10^{19}$ |
| $\alpha_H$ | | $9 \cdot 10^{-4}$ | $1.1 \cdot 10^{-3}$ | $1.8 \cdot 10^{-3}$ | $2.1 \cdot 10^{-3}$ |
| $S_{\Delta R}{}^2$ | Ω²/Hz | $1 \cdot 10^{-3}$ | $1 \cdot 10^{-2}$ | - | - |

## IV. Model validation

The LFN parameters extracted in Section III are used in the simulations of the Verilog-A model [18]-[19] and the results are shown in right column of Fig. 5, 6 for the GFETs available. Measured LFN $S_{Ibf}$ at *1 Hz* is depicted with markers, total model [19] with solid lines and model without VS effect with dashed lines [18] for C300 GFET in right-upper plot of Fig. 5 and D100 GFET in right-bottom plot of Fig. 5, respectively. $S_{Ibf}$ is shown vs. $V_{DS}$ in left subplots and vs. $I_D{}^2$ in right subplots for $V_G = 0.7$, $0.8$, $0.9$ V and the evaluation of the model is very consistent. It is remarkable that $S_{Ibf}$ follows a $\sim V_{DS}{}^2$ and a $\sim I_D{}^2$ dependence up to the highest measured $V_{DS}$ value for each device which resembles the output characteristic behaviour at linear region, which is the case for the specific operating conditions. It can be easily concluded, especially for D100 device where VS effect is more intense, that the reduction of $S_{Ibf}$ at higher drain voltage decreases the slope of the dependences mentioned above. If VS effect is ignored, then $S_{Ibf}$ follows $V_{DS}{}^2$ and $I_D{}^2$ lines precisely. In order to examine the $V_{DS}$ dependence of $S_{Ibf}$ thoroughly and confirm the contribution of VS effect, the model is extended to higher $V_{DS}$ values up to *1 V* for $V_{GS} = 1.2$ V (This gate voltage is chosen in order to remain near CNP) for the D100 GFET. As it is clear from right-bottom plot of Fig. 5, total $S_{Ibf}$ model starts to get saturated at higher $V_{DS}$, without following $\sim V_{DS}{}^2$ and $I_D{}^2$ behaviour anymore while the long-channel model, where VS effect is not included, retains this behaviour. In right plot of Fig. 6, $S_{Ibf}$ vs. $V_{GEFF}$ is shown for the SG-GFET and the model is also precise.

## V. Conclusions

In conclusion, a thorough LFN parameter extraction methodology is proposed for SL GFETs. The $\Delta\mu$, $\Delta N$ and $\Delta R$ effects that contribute to LFN are analyzed and the relevant parameters are derived from the $V_{GEFF}$ regions that they prevail for low $V_{DS}$ values. At high $V_{DS}$ region, the reduction of LFN due to VS effect is accurately predicted by the model and the VS related IV parameter is confirmed. The extracted parameters are applied at the recently established LFN GFET model [18]-[19] which gives excellent results when validated with experimental data from SL short-channel CVD GFETs and long channel SG-GFETs for a wide range of bias conditions. This work can be proved a very reliable tool for RF graphene circuits' design where LFN is significant and thus, should be accurately predicted.

## Acknowledgements

This work was funded by the European Union's Horizon 2020 research and innovation program under Grant Agreement No. GrapheneCore2 785219 (Graphene Flagship), Marie Skłodowska-Curie Grant Agreement No 665919 and Grant Agreement No. 732032 (BrainCom). We also acknowledge financial support by Spanish government under the projects TEC2015-67462-C2-1-R, RTI2018-097876-B-C21 (MCIU/AEI/FEDER, UE) and 001-P-001702 (RIS3CAT). This work was partly supported by the French RENATECH network. The ICN2 is also supported by the Severo Ochoa Centres of Excellence programme, funded by the Spanish Research Agency (AEI, grant no. SEV-2017-0706).